# Tunable Magnetism and Insulator-Metal Transition in Bilayer Perovskites

Shaowen Xu[1*], Fanhao Jia[1*], Guodong Zhao[1], Tao Hu[1], Shunbo Hu[1], and Wei Ren[1†]

[1]*Physics Department, Shanghai Key Laboratory of High Temperature Superconductors, State Key Laboratory of Advanced Special Steel, International Centre of Quantum and Molecular Structures, Shanghai University, Shanghai 200444, China*

[†]renwei@shu.edu.cn

*These authors contributed equally to this work.

**Abstract** Two-dimensional (2D) transition-metal oxide perovskites greatly expand the field of available 2D multifunctional material systems. Here, based on density functional theory calculations, we predicted the presence of ferromagnetism orders accompanying with an insulator-metal phase transition in bilayer $KNbO_3$ and $KTaO_3$ by applying strain engineering and/or external electric field. Our results will contribute to the applications of few-layer transition metal oxide perovskites in the emerging spintronics and straintronics.

Freestanding transition-metal oxide perovskites have been successfully synthesized, which opened a promising door to design applications in low-dimensional multifunctional electronic devices [1-6]. Compared to conventional 2D materials, such as graphene or transition-metal dichalcogenides [7-9], freestanding perovskite systems are believed to present enriched functionalities from the strongly correlated states in accompany with fascinating oxygen octahedron distortions [10-16]. Recently, oxide perovskites membranes have been demonstrated to exhibit giant tetragonality, flexibility, polarization [1], piezoelectricity [3] and super-elasticity [2]. Most of the interest was mainly focused on properties related to mechanics under the quantum confinement, leaving their electronic and magnetic properties with definitely similar fundamental scientific importance largely unexplored.

Bulk $KNbO_3$ (KNO) and $KTaO_3$ (KTO) perovskites have been studied for over 70 years [17]. The former is famous as room-temperature lead-free ferroelectric material [18], while the latter is so-called incipient ferroelectric [19]. Some exotic properties based on them have been realized by doping, size, dimension or interface engineering, including the magnetic response in nanocrystalline KNO, two-dimensional electron gas

(2DEG) in their various heterostructures [20,21], 2DEG in KTO polar surface with strong spin-orbit coupling [22,23], and the superconductivity in KTO by electrostatic carrier doping [24]. It is thus expected that their electronic and magnetic properties will be strongly modulated when the perovskites are approaching the 2D limit.

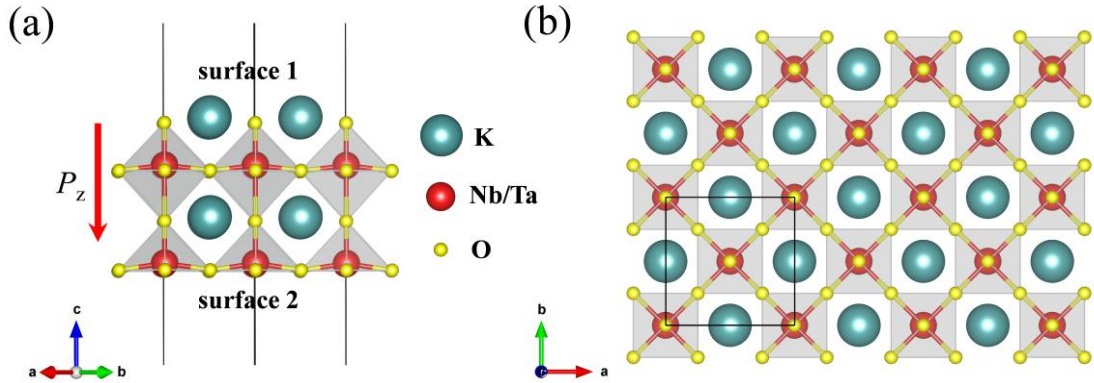

Figure. 1 The schematic (a) side and (b) top views of crystal structure of bilayer KTO or KNO.

To determine theoretically the ground-state structure and properties of bilayer KNO and KTO, we performed density functional theory (DFT) calculations within the generalized gradient approximation (GGA) in the form proposed by Perdew, Burke, and Ernzerhof (PBE) [25] as implemented in the Vienna *ab initio* simulation package (VASP) [26]. In the calculations of electronic and magnetic properties, an effective Hubbard parameter $U_{\text{eff}}$ was adopted for the Nb (3.6 eV) $4d$ and Ta (3.0 eV) and $5d$ states [27]. The hybrid functional approach HSE06 [28] was also applied to calculate the band gaps. The energy cutoff was chosen to be 600 eV. The 12×12×1 and 20×20×1 Γ centered Monkhorst-Pack k-grids were used for √2×√2×1 cell and cubic unit cell, respectively. The convergence criterion of energy was set to at least smaller than $10^{-7}$ eV and atomic positions were fully relaxed until the maximum force on each atom was less than $10^{-3}$ eV/Å. After adding a more than 20 Å vacuum space into their cubic unit cell, we additionally used a √2×√2×1 supercell to allow for in-plane antiferrodistortive oxygen motions. Starting from this supercell corresponding to the highest achievable symmetry, we identified instabilities from the phonon dispersion spectrum, then accordingly lowered the symmetry and performed new structural relaxations. The phonon dispersion was calculated by density-functional perturbation theory (DFPT) method as implemented in the PHONOPY package [29,30]. This process was repeated until the instabilities were minimized (see Figure S1). At last, we found their ground-state structures do adopt a √2×√2×1 cell maintaining a mirror symmetry plane that is perpendicular to the b axis (see Figure 1), giving lattice parameters a = 5.6028 (5.5745) and b = 5.6027 (5.5744) Å for bilayer KNO (KTO). The extremely small difference of in-plane lattice constants ($10^{-4}$ Å) is accompanied with ferroelectric displacements along the a-axis or the pseudocubic (ps) $[110]_{ps}$ direction, which results double-well-like potential curves with negligible well depths ΔE to be ~0.01 meV. These in-plane polarizations are intrinsically tiny due to the suppression by the more dominant out-of-plane polarizations. It is easy to find that these systems naturally show an out-of-plane symmetry breaking. Both bilayer systems show a relatively large tetragonality c/a ratio ~1.10 (see Table I) which is even comparable with the bulk value in $PbTiO_3$ [31].

Table I. The PBE optimized lattice constants $a$ (Å) of bilayer (BL) KNO and KTO based on √2×√2×1 cells and experimental values of their bulk cubic phases. The tetragonality c/a is defined as the ratio between the out-of-plane and in-plane distances of $K^+$ ions. The planar averaged electrostatic potential differences between the two sides of the slab along the z-direction Δφ (eV) are also presented. The out-of-plane polarization $P_z$ ($\mu C/cm^2$) was calculated from the integral of charge density. The energy gaps (eV) of two BL systems were calculated by using HSE06.

| System | $a$ | c/a | $P_z$ | Δφ | Gap |
|---|---|---|---|---|---|
| BL KTO | 5.575 | 1.10 | -2.00 | -2.62 | 3.21 |
| BL KNO | 5.603 | 1.10 | -2.04 | -2.96 | 2.32 |
| $KTO_{cubic}$ | 3.988[a] | / | / | / | 3.64[c] |
| $KNO_{cubic}$ | 4.022[b] | / | / | / | 3.24[d] |

[a] [32], [b] [33], [c] [34], [d] [35]

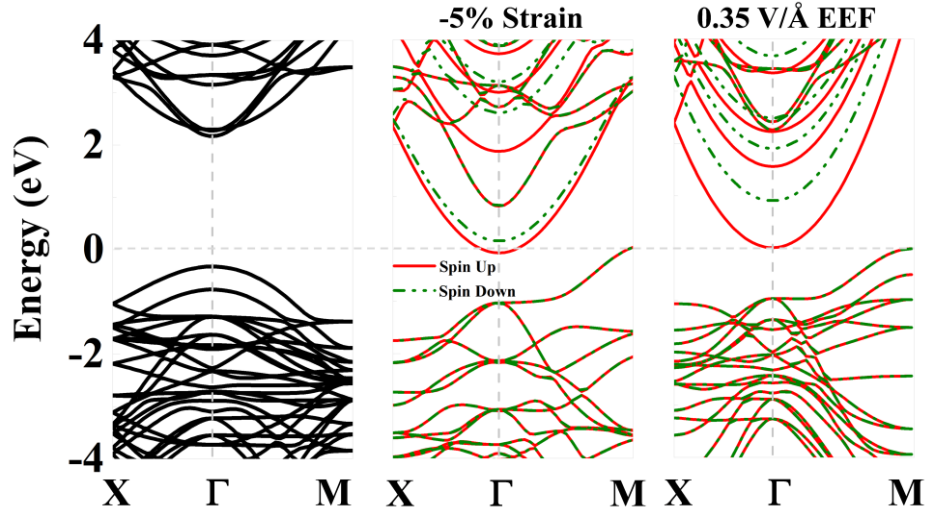

Figure. 2 The PBE+$U$ band structures of BL KTO for (a) the ground state, (b) under a biaxial strain of -5%, and (c) under 0.35V/Å external electric field. The BL KNO has similar response to the biaxial strain or the electric field.

We displayed the PBE+$U$ band structure of ground state BL KTO in Figure 2(a). Due to the tiny in-plane displacement along the $[110]_{ps}$ direction, it shows a direct band gap with both the valence band maximum (VBM) and conduction band minimum (CBM) located at the Γ point. The calculated band gap by using HSE06 is 3.21 eV for BL KTO, and 2.32 eV for BL KNO, which are smaller than their experimental bulk cubic values. The VBM is majorly derived from O $2p_x$ with a slight hybridization of Ta (Nb) $5d$ ($4d$) states, while the CBM is mainly derived from Ta (Nb) $5d$ ($4d$) states (Figure S2). The electron effective mass of BL KTO is 0.31 (0.45) $m_0$ along Γ-X (Γ-M) direction, and the hole effective mass is -1.70 (-1.17) $m_0$ along Γ-X (Γ-M) direction. These values are slightly larger than those of the BL KNO when no stain or field is applied.

The strain engineering and external electric field (EEF) have been extensively considered to tune physical properties in various layered materials [36-40], derived from the stress or interface charge transfer of substrate which are ubiquitous in the experiment. These effects are also possible to change electron ordering and orbital occupancy which may induce magnetism in nonmagnetic systems. Intrinsic bilayer KTO and KNO are nonmagnetic, but our calculations show that they might transform to ferromagnetic metal states under a compressive biaxial strain or under a high EEF.

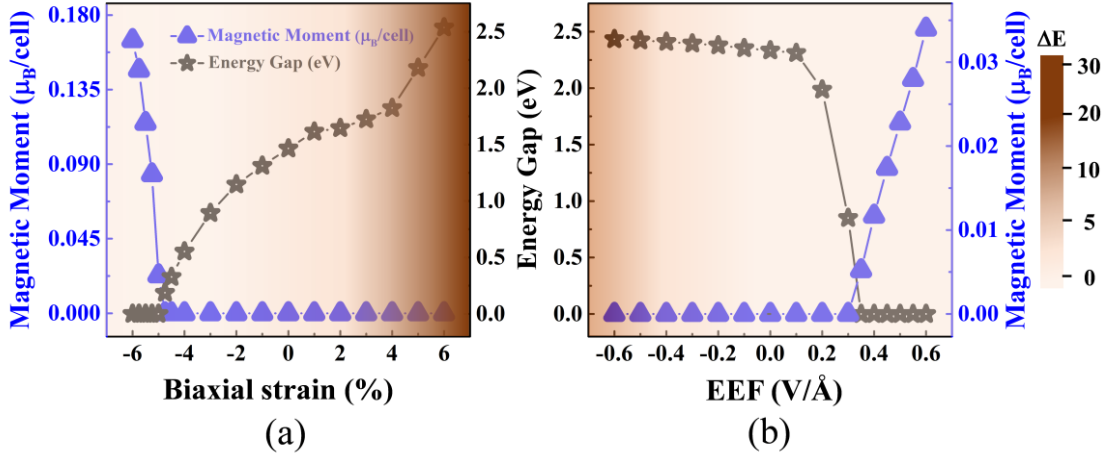

Figure. 3 The PBE+$U$ band gap (stars) and magnetic moment (triangles) of BL KTO as functions of (a) biaxial strain and (b) EEF intensity. The background color represents the double well depth $\Delta E$ (meV/cell) which is defined by the energy difference between the structures with in-plane displacements and without in-plane displacements.

We studied the electronic and magnetic properties of BL KTO and KNO under biaxial strain from -6 % to 6 %. The strain is defined as $\varepsilon(\%) = \frac{a-a_0}{a_0}$, where $a$ and $a_0$ are the strained and equilibrium lattice constants, respectively. It is worth mentioning that recent experiments have realized a maximum strain of ~10% in the BTO nanobelt without any fracture [2]. The large tunability of epitaxial strain in ferroelectricity and electronic properties has been explored in oxide perovskite (e.g. $SrTiO_3$) membranes [41] or their heterostructures [42]. As shown in Figure 3(a) (see Figure S3 for BL KNO), the tensile strain increases the in-plane ferroelectric double well depth $\Delta E$ dramatically from 0.01 meV at zero strain to 30 meV at 6% tensile strain. The tensile strain also monotonically increases the band gap of BL KTO, while the compressive strain may effectively suppress the band gap down to zero at ~ -5%.

It is possible to induce an insulator-metal transition at ~ -5% compressive strain as we displayed the band structure in Figure 2(b). During the evolution of this transition, we have found that there are two interesting observations. The first effect is that the in-plane ferroelectric displacements are fully eliminated, as a result the unit cell is strictly reduced to a 1×1×1 primitive cell where the hole pocket of Fermi surface is located at

M point rather than at Γ point. The second effect is the previous empty $d$ states of the conduction band now become partially occupied, giving rise to spin split and polarized states crossing the Fermi level. Furthermore, the magnetic moment is found to be linearly increased if the stronger compressive strains are applied beyond -5%. The amount of occupied $d$ states is also enhanced gradually by the compressive strains, according to the analysis of projected density of states (see Figure S5). As a matter of fact, the large density of states $D(E_f)$ accumulated at the Fermi level ($E_f$) under highly compressive strains is the origin of ferromagnetism which can be explained by the Stoner model [43]. More importantly, we found the spin density is mainly accumulated at the KO$^-$ surface 1 which forms a spin polarized 2DEG, as displayed in Figure S4. Accordingly, a 2D hole gas (2DHG) was also found to be accumulated at the TaO$_2^+$ surface 2. Such a pair of 2DEG and 2DHG states have also been found in the KTO thicker slabs in the vacuum when their thicknesses were larger than 7 unit cell. Interestingly these surface 2D electron and hole gases can be tuned by applying biaxial strain, i.e. the compressive in-plane strain decreases the 2D carrier concentrations down to zero and a new pair of surface 2D electron and hole gases appear at the reversed surfaces[27]. Here we reveal that when the insulator-metal transition occurs, the conduction bands approaching the Fermi level become spin-polarized. This explains why we could induce the magnetic moments through the epitaxial strain or electric field effect. The Bader charge analysis (Figure S6) can further give quantitatively the tendency of the charge transfer between atomic layers with increasing compressive strain.

EEF is another very efficient way to turn the electronic and magnetic properties of 2D materials. Here, by the positive EEF intensity we mean that the direction of the applied electric field is along +$c$ direction, and vice versa. The EEF effects on structural, electronic, and magnetic properties are displayed in Figure 3(b), Figure S7 and S8. Firstly, the in-plane double-well energy barrier Δ$E$ can be enhanced by applying a negative EEF which can also promote the magnitude of out-of-plane polarization $P_z$. As listed in Figure S9, a positive EEF is opposite to the polarization field, as a result, it will suppress the absolute value of $P_z$. However, when the EEF is larger than 0.3 V/Å, the polarization $P_z$ will be flipped to the opposite direction. Then, the positive EEF turns to increase the $P_z$ approximately linearly.

Now we look at the planar averaged electrostatic potential difference $\Delta\varphi$ between the two sides of the slab along the *c*-direction. The $\Delta\varphi$ is proportional in magnitude to the polarization $P_z$ which is perpendicular to the slab plane. Under the zero-EEF, the $\Delta\varphi$ of -2.62 eV represents the intrinsic built-in electric field, which means that the direction is along the -*c* axis. Remarkably, the $\Delta\varphi$ can be linearly changed from -10.0 eV to 7.5 eV by the EEF ranged from + 0.5 to - 0.5 V/Å. The widely tunable $\Delta\varphi$ enables the KTO bilayer to be a desirable candidate for field emission devices [44].

We found the negative EEF has very little effect on the band gap, but it will dramatically reduce the band gap if a positive EEF larger than 0.2 V/Å is applied. The band structure of bilayer KTO under a 0.35 V/Å EEF was displayed in Figure 2c., which was the critical point for a transition from spin-degenerate insulator to ferromagnetic metal. Moreover, the two accompanied effects previously mentioned at large compressive strain were also existing under high positive EEF.

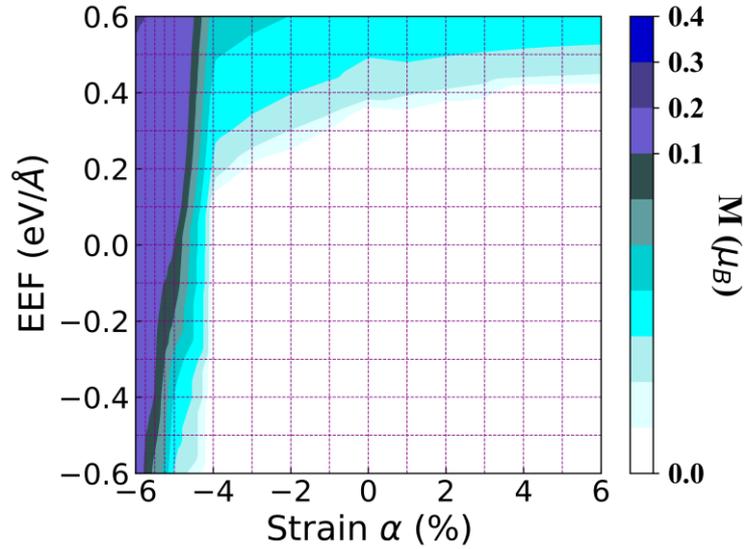

Figure 4. Phase diagram of the magnetic moment versus the strained bilayer KTO with different external electric fields.

We also examined the combined effect of biaxial strain and EEF to figure out how they could promote each other. The phase diagram of the magnetic moment of BL KTO with respect to the strain and EEF is displayed in Figure 4 (see Figure S11 for BL KNO). We observe that the compressive strain and the positive EEF can increase the magnetic moment jointly. The origin of such magnetization control is from the fact that they both have the similar effects on the electrostatic potential difference $\Delta\varphi$. More specifically, for example, under a -3% strain combined with a 0.25 V/Å EEF the bilayer KTO reaches the ferromagnetic metal state. This can be attributed to the positive EEF induces an additional shift of Fermi level $E_F$ to allow more unoccupied $d$ states to be occupied in the strained systems (see the band structures in Figure S12), which results in more pronounced magnetic moment. On the other hand, the strain engineering tends to result in larger magnetic moment than the EEF, possibly due to the larger structure distortion. If realized in experiment, such tunability of magnetic moment is crucially beneficial for applications in spintronics and straintronics.

In conclusion, we theoretically studied the biaxial strain and EEF effects on the 2D bilayer $KTaO_3$ and $KNbO_3$. A transition from a non-magnetic insulator to a ferromagnetic metal might be realized by applying compressive strain and/or positive EEF. The structural details with in-plane displacements, out-of-plane polarization, and electrostatic potential difference $\Delta\varphi$ were systematically investigated from first principles. The predicted effects reported in this work will be useful to guide future experiments on the 2D perovskites.

## ACKNOWLEDGMENTS

This work was supported by the National Natural Science Foundation of China (Nos. 51672171, 51861145315 and 51911530124), Shanghai Municipal Science and Technology Commission Program (No.19010500500), State Key Laboratory of Solidification Processing in NWPU (SKLSP201703), Austrian Research Promotion Agency (FFG, Grant No. 870024, project acronym MagnifiSens), and Independent Research Project of State Key Laboratory of Advanced Special Steel and Shanghai Key Laboratory of Advanced Ferrometallurgy at Shanghai University. F.J. is grateful for the support from the China Scholarship Council (CSC).